\documentclass[doublecol,figures]{epl2}          
\usepackage{amssymb}                           
\title{Ion pump activity generates fluctuating electrostatic forces in biomembranes}                                   
\shorttitle{Ion pump activity} %Insert here a short version of the title if it exceeds 70 characters                  
\author{B. Loubet\inst{1} \and M. A. Lomholt\inst{1}}  
\shortauthor{B. Loubet \etal}                 
\institute{
  \inst{1} MEMPHYS - Center for Biomembrane Physics, Department of Physics and Chemistry, University of Southern Denmark, Campusvej 55, 5230 Odense M, Denmark}

\abstract{
We study the non-equilibrium dynamics of lipid membranes with proteins that actively pump ions across the membrane. We find that the activity leads to a fluctuating force distribution due to electrostatic interactions arising from variation in dielectric constant across the membrane. By applying a multipole expansion we find effects on both the tension and bending rigidity dominated parts of the membranes fluctuation spectrum. We discuss how our model compares with previous studies of force-multipole models.                           
}

\newcommand{\e}[0]{\epsilon}
\newcommand{\dd}[0]{\mathrm{d}}

\newcommand{\Eqref}[1]{eq.~(\ref{#1})}
\newcommand{\EQref}[1]{Eq.~(\ref{#1})}
\newcommand{\Figref}[1]{fig.~\ref{#1}}

\newcommand{\wvec}{\vect{w}}

\usepackage{amsmath}

\begin{document}

\maketitle

\section{Introduction}
Biomembranes are self-assembled structures containing lipids and proteins that
form selective barriers surrounding cells and organelles. They also participate
actively in many biological processes, including creation and maintenance of   
ion gradients through the dissipation of free energy \cite{alberts02}. The mechanical properties of a fluid lipid-protein membrane in thermal equilibrium have successfully been described
based on the Helfrich Hamiltonian \cite{helfrich73}, which include quantities such as bending rigidity and tension, and for asymmetric membranes also spontaneous curvature. From this starting point observed properties of both membrane shapes as well as their thermal fluctuations can be derived \cite{seifert97}. The situation, however, becomes different when the membrane is driven out of equilibrium by some active process. For these active membranes a clear general picture has not yet emerged that explains their mechanical properties. 

That non-equilibrium activity can significantly alter mechanical properties has been demonstrated in experiments involving ion pumps in lipid membranes. These experiments include mechanical manipulations by micropipettes \cite{manneville99,girard05} as well as observations of shape fluctuations by video microscopy \cite{faris09}. The apparent interpretation of the experiments is that the activity tends to enhance fluctuations of the membrane
shape. To explain this enhancement it was proposed in \cite{manneville01} that
an effect of the ion pump activity is to induce a force-multipole in the
membrane environment. By Newton's 3rd law a monopole is excluded, so a shifted
force-dipole was chosen as the simplest possibility allowed by symmetry.
This force-dipole model was further studied in
\cite{sankararaman02,lacoste05,lomholt06,lomholt06b}.

A shortcoming of \cite{manneville01,sankararaman02,lacoste05,lomholt06,lomholt06b} is that no specific physical explanation is given for the origin of the force-dipoles. In this letter we propose that these forces can arise due to electrostatic interactions when an ion is transported across the low dielectric membrane interior between the surrounding highly polariseable water. In the following we calculate electrostatic forces and consequences for fluctuations within a simple model of this process and compare with previous work on the force-dipole model. We find that our model agrees with a force-multipole picture of the activity. However, contrary to the assumptions in the previous work \cite{manneville01,sankararaman02,lacoste05,lomholt06,lomholt06b}, we find that our model leads to a symmetric coupling between pump diffusion and membrane shape dynamics.

\section{Electrostatic model}
\begin{figure}
\onefigure{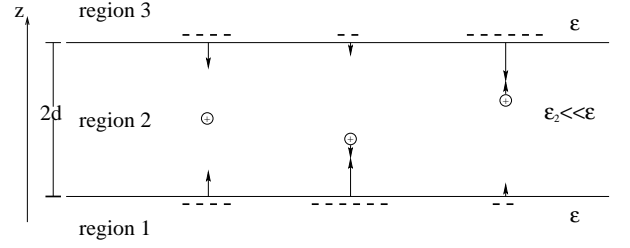}
\caption{The lipid bilayer is portrayed as a band of dielectric of permittivity $\e_2$ embedded inside a dielectric of permittivity $\e$. The dielectric media polarisation, due to the presence of the positive ions, give rise to a net negative charge at the interfaces. These charges produce forces on the membrane interfaces and on the ions.}                          
\label{fig:MembSchema}
\end{figure}

Our model for the membrane is a surface of thickness $2 d$ with a dielectric
permittivity $\e_2$ embedded in a fluid of dielectric permittivity $\e$, see
\Figref{fig:MembSchema}. We choose Cartesian coordinates $(x,y,z)$ with a basis of unit vectors $\vect{\hat{x}}$,
$\vect{\hat{y}}$, $\vect{\hat{z}}$. Taking the $z$-axis along the normal direction to the membrane we call the fluid region below the membrane, $z<-d$, region 1. The membrane region, $-d<z<d$, is region 2 and the fluid at $z>d$ is region 3. 
In the membrane region 2 there is a charge $q_0$ situated at $(x_c,y_c,z_c)$, which is in the process of being pumped across the membrane. We are interested in deriving the forces in our system due to the presence of the charge.             
To calculate these forces we use the method of image charges \cite{jackson98}. By this we obtain the potentials $\phi_i$ in each region labelled by $i$:
\begin{align}
        \phi_1 (x,y,z) & = \frac{b}{4 \pi \e} \sum_{n=0}^{\infty} a^n \frac{q_0}{r_3^{(n)}}\\
\label{phi2}
        \phi_2 (x,y,z) & = \frac{q_0}{4 \pi \e_2} \left[ \frac{1}{r^{(0)}} + \sum_{n=1}^{\infty} a^n  \left( \frac{1}{r_1^{(n)}} + \frac{1}{r_3^{(n)}} \right)  \right]\\
        \phi_3 (x,y,z) & = \frac{b}{4 \pi \e} \sum_{n=0}^{\infty} a^n \frac{q_0}{r_1^{(n)}}
\end{align}
where $a = (\e_2 -\e )/(\e_2 +\e)$, $b = 2 \e/(\e_2 +\e)$, and $r_1^{(n)}$ and $r_3^{(n)}$ are the distances from the $n^\mathrm{th}$ image charge in regions 1 and 3 to the point $(x,y,z)$:
\begin{equation}
        r_i^{(n)} = \sqrt{ ( x - x_c )^2 + (y - y_c)^2 + (z- z_i^{(n)})^2}
\end{equation}
with $i=1,3$, $z_1^{(n)} = (-1)^n z_c - 2 n d$, and $z_3^{(n)} = (-1)^n z_c + 2 n d$. Finally $r^{(0)} =r_i^{(0)}$ is the distance to the actual charge situated in the membrane.
These potentials give rise to forces pointing along the normal direction to the membrane at three locations: two forces distributed on the membrane-fluid interface areas due to discontinuity of the Maxwell stresses
\begin{align}
        f_{1,2}(x,y) & = T_{2,zz}(x,y,-d) - T_{1,zz}(x,y,-d)\\
        f_{2,3}(x,y) & = T_{3,zz}(x,y,d) - T_{2,zz}(x,y,d)
\end{align}
where the Maxwell stresses along the normal direction is
\begin{equation}
T_{i,zz}(x,y,z) = \frac{\e_i}{2} \left( \left( \frac{\partial \phi_i}{\partial z} \right)^2 - \left(  \frac{\partial \phi_i}{\partial x} \right)^2 - \left( \frac{\partial \phi_i}{\partial y} \right)^2  \right)\label{StressTensor}            
\end{equation} 
and $\e_i=\e$ for $i=1,3$. The third force arises from the image charges interactions with the real ion:
\begin{align} 
 f_{\rm ion}(x_c,y_c,z_c) & = - q_0 \partial_z [ \phi_2 - q_0/(4\pi \e_2 r^{(0)})]\nonumber\\
& = \frac{q_0^2}{4 \pi \e_2} \sum_{n=1}^{\infty} \left[ \frac{a^n}{ ( z_c - z_1^{(n)} )^2 } - \frac{a^n}{(  z_c - z_3^{(n)}  )^2 }\right].
\end{align}                                                                                     
Because of continuity of the lateral components of the electric field across the membrane interfaces there are no lateral components to the forces.                                                            
The next step in our derivation is to approximate the interface forces $f_{1,2}(x,y)$ and $f_{2,3}(x,y)$ with point forces at the interface. Systematically this approximation corresponds to expanding the Fourier transforms of these forces to zeroth order in powers of $d$ times wavenumber $q$. With the convention that the Fourier transform of a function $g(x,y)$ is denoted with a bar
\begin{equation}
        \bar{g}(q_x,q_y) = \int_{- \infty}^{+ \infty} \dd x \int_{- \infty}^{+ \infty} \dd y \ e^{-i(q_x x +q_y y)} g(x,y)
\end{equation}
and $q = \sqrt{q_x^2 +q_y^2}$ we obtain
\begin{align}                         
f_{1,2}(x,y) & \approx \delta(x-x_c)\delta(y -y_c)\tilde{f}_{1,2}(z_c)\\
f_{2,3}(x,y) & \approx \delta(x-x_c)\delta(y -y_c)\tilde{f}_{2,3}(z_c)
\end{align}                                                          
where                                               
\begin{align}                         
\label{force1}                       
\tilde{f}_{1,2}(z_c) & = {\bar f}_{1,2}|_{q=0}= \frac{- q_0^2}{4 \pi \e_2} \sum_{n=1}^\infty \sum_{m=0}^\infty \frac{a^{n+m}}{\left( z_3^{(m)} - z_1^{(n)} \right)^2 }\\  
\label{force3} 
\tilde{f}_{2,3}(z_c) & = {\bar f}_{2,3}|_{q=0}= \frac{q_0^2}{4 \pi \e_2} \sum_{n=1}^\infty \sum_{m=0}^\infty \frac{a^{n+m}}{\left( z_3^{(n)} - z_1^{(m)} \right)^2 }                                
\end{align}                                                           
Combining the three contributions for all pumps that are in the process of transporting an ion across the membrane we find the total force arising due to the activity                                 
\begin{multline}
        \vect {F}_{\rm act}(x,y,z) = \sum_{n=1}^{N_p} \delta(x-x_{c,n})\delta(y-y_{c,n})
 \Big[\delta(z-z_{c,n})\\                                                              
\times f_{\rm ion}(z_{c,n}) 
 + \delta(z+d) \tilde{f}_{1,2}(z_{c,n}) + \delta(z-d) \tilde{f}_{2,3}(z_{c,n}) \Big] \vect{\hat{z}}\label{eq14}
\end{multline}
where $(x_{c,n},y_{c,n},z_{c,n})$ are the coordinates of the $n^{\rm th}$ charge in the process of going through the membrane.
We are here assuming a sufficiently low density of pumps such that we can neglect interactions between the different ions.   
\EQref{eq14} is one of the central results of this paper. It shows that                             
our model to a good approximation falls within the class of force-multipole models for membrane activity. We can therefore use the more general results of \cite{lomholt06} to obtain the consequences of the activity for the dynamics of the membrane shape.             

\section{Net active force on the membrane surface}

In the following we will parametrize
the membrane shape by the height $z=h(x,y,t)$ of its midplane above the $xy$-plane at time $t$. The equation of motion for the membrane shape $h$ will be the normal component of the membrane force balance equation. Assuming that we are at low Reynolds number with most of the dissipation occurring in the surrounding water we can write the force balance per area as 
\begin{equation} 
\label{forcebalance}
        \vect{f}_{\rm rs} + \vect{T}^+ + \vect{T}^- + \vect{f}_{\rm act} = 0  
\end{equation}                            
where $\vect{f}_{\rm rs}$ is the restoring force associated with elastic properties of the membrane, $\vect{T}^+$ and $\vect{T}^-$ are the stresses from the bulk fluid on each side of the membrane, and $\vect{f}_{\rm act}$ is the additional force induced by the activity. We will derive equations of motion for the membrane to first order in deviations from a uniform flat membrane. Since by symmetry the lateral components of the forces must vanish at zeroth order, we are allowed to simply identify the normal component of the forces with the $z$ components in the following (the difference being of higher order in the deviations).

The derivation of $\vect{f}_{\rm act}$ starting from a force distribution of the form given by \Eqref{eq14} was accomplished perturbatively by a moment expansion in \cite{lomholt06}. This derivation was achieved by studying the lateral stresses and bending moments induced in the membrane by the force distribution, and then using that these quantities can be related to the membrane force balance \cite{evans80,kralchevsky94,lomholt06c}. The end result up to the second moment for this calculation for a nearly planar membrane is \cite{lomholt06}  
\begin{equation} 
\label{eq16}  
\vect{f}_{\mathrm{act}}\cdot\vect{\hat{z}} = \sigma_{\rm dip} \Delta h + \frac{1}{2} \Delta Q, 
\end{equation}
where $\Delta=\partial^2/\partial x^2+\partial^2/\partial y^2$ is the 2D Laplacian and $\sigma_{\rm dip}$ and $Q$ are the first (dipole) and second (quadrupole) moment of the force distribution  
\begin{align}  
        \sigma_{\mathrm{dip}} &= \int \dd z \ z\vect{F}_{\rm act}(x,y,z)\cdot \vect{\hat{z}}\\  
        Q &= \int \dd z \ z^2 \vect{F}_{\rm act}(x,y,z)\cdot\vect{\hat{z}}  
\end{align} 
Apart from the unitless numerical prefactors then one could also have written down \Eqref{eq16} based on \Eqref{eq14} using symmetry arguments, linearity and dimensional analysis.  

Based on the force distribution (\ref{eq14}) we will expand the moments as \cite{REM1}
\begin{align}\label{eq19} 
\sigma_\mathrm{dip} &= n_\Sigma^0 F_{(1)}\\  
 Q & =  n_\Delta F_{(2)} + S\label{eq20}   
\end{align}        
where $n_\Sigma^0$ is the average concentration of pumps per area, $n_\Delta=n_\Delta(x,y,t)$ is the local concentration difference between pumps transporting ions in the positive and negative $z$-direction, and
\begin{align}   
\label{firstmoment} 
        F_{(1)} & = \int \dd z_c \ \lambda(z_c) \left[ -d \tilde{f}_{1,2}(z_c) + d \tilde{f}_{2,3}(z_c) + z_c f_{\rm ion}(z_c)  \right] \\      
\label{secondmoment}  
        F_{(2)} & = \int \dd z_c \ \lambda(z_c)  \left[ d^2 \tilde{f}_{1,2}(z_c) + d^2 \tilde{f}_{2,3}(z_c) + z_c^2 f_{\rm ion}(z_c)  \right]      
\end{align}    
are averaged moments for a single pump with distribution $\lambda(z_c)$ for the ions position in the membrane for a positively oriented pump (the negatively oriented having a reflected distribution. This reflection is the reason behind the occurrence of the difference $n_\Delta$ in \Eqref{eq20}). We will assume that the overall membrane system is symmetric under reflection of the $z$-direction such that the average value of $n_\Delta$ vanishes (up-down symmetry). Finally, $S$ is a zero-average noise term. We have neglected noise in $\sigma_{\rm dip}$, since $\sigma_{\rm dip}$ enters a term in eq. (\ref{eq16}) which is already first order in deviations from uniformity. Inserting eq. (\ref{eq19}) and (\ref{eq20}) in eq. (\ref{eq16}) we arrive at  
\begin{equation} 
\label{activeforce}
        \vect{f}_{\mathrm{act}}\cdot\vect{\hat{z}} = n_\Sigma^0 F_{(1)} \Delta h + \frac{1}{2} F_{(2)} \Delta n_\Delta + \frac{1}{2} \Delta S.    
\end{equation} 

\section{Dynamics and fluctuations}

A closed set of equations governing the dynamics of $h$ can now be obtained by specifying the remaining forces and a dynamic equation for $n_\Delta$. The restoring force $\vect{f}_{\rm rs}$ can be obtained by functionally differentiating the free energy $F$ of the membrane \cite{lomholt05}. Supplementing the Helfrich free energy \cite{helfrich73} with coupling to the $n_\Delta$ field we can take $F$ to be  
\begin{equation} 
\label{freeenergy}
        F = \int \dd A \left( \frac{\kappa}{2} \left( 2 H \right)^2 + \sigma_0 + \Lambda n_\Delta H + \frac{\chi}{2}n_\Delta^2  \right) , 
\end{equation} 
where $H\approx \Delta h/2$ is the mean curvature, $dA\approx dxdy\{1+[(\partial h/\partial x)^2+(\partial h/\partial y)^2]/2\}$ the area measure, $\kappa$ the bending rigidity, $\sigma_0$ a tension, $\Lambda$ a coupling constant, and the parameter $\chi\approx k_B T/n_\Sigma^0$ for sufficiently dilute pump density due to the entropy of mixing. Linear terms in $n_\Delta$ and $h$ are excluded by up-down symmetry. 
Functionally differentiating with respect to $h$ gives the restoring force
\begin{equation} 
\label{restoringforce} 
        \vect{f}_{rs}\cdot\vect{\hat{z}}= -\frac{\delta F}{\delta h} =  \sigma_0 \Delta h - \kappa \Delta^2 h - \frac{\Lambda}{2} \Delta n_\Delta + O(h^2).
\end{equation} 
The motion of the bulk water surrounding a lipid vesicle will be governed by the low Reynolds number Navier-Stokes equation. From a no-slip boundary condition where the water and membrane velocity are matched one can obtain the stress that the water exert on the membrane. In Fourier space this leads to the force \cite{lomholt06}   
\begin{equation}
        \left( \bar{\vect{T}}^+ + \bar{\vect{T}}^- \right)\cdot\vect{\hat{z}} = - 4 \eta q \dot{\bar{h}} + {\bar \eta}_{\rm hydro}  
\end{equation} 
where
the bar denotes the Fourier transform, 
the dot denotes a time derivative,
$\eta$ is the bulk viscosity and $\eta_{\rm hydro}$ is the thermal noise. The strength of this noise can be obtained by applying the fluctuation-dissipation theorem.                   

The dynamic equation for the pump density difference is a continuity equation. We will write it as
\begin{equation}
\label{diffusion} 
        \dot{n}_\Delta = - \vect{\nabla}_{\|} \cdot \left\{-\Omega \vect{\nabla}_{\|} \left[ \frac{1}{2} \Lambda_a \Delta h + \chi_a n_\Delta \right] + \vect{\eta}_{\|,{\rm diff}}\right\},
\end{equation} 
where $\vect{\nabla}_{\|}$ is the gradient operator restricted to the membrane (and thus $\vect{\nabla}_{\|}\cdot\vect{\nabla}_{\|}=\Delta$), $\Omega$ is a phenomenological transport coefficient and $\vect{\eta}_{\|,{\rm diff}}$ is the corresponding thermal noise. The term in the square brackets of \Eqref{diffusion} is the functional derivative of $F$ with respect to $n_{\Delta}$, except that the index $a$ indicates that $\Lambda$ and $\chi$ can be modified by the activity. That the collective diffusion constant $D_a\equiv \Omega\chi_a$ can change upon pump activation has been observed experimentally for the proton pump bacteriorhodopsin \cite{kahya02}. 
In the following we will discuss how a change of $\Lambda$ is related to the active force distribution.  

We assume that the membrane is situated in a quadratic frame of area $L^2$ with periodic boundary conditions. This leads to a decoupling of the dynamics of the different Fourier modes of $h$ and $n_\Delta$. Introducing the column vector $\wvec=({\bar h}(q,t),{\bar n}_\Delta(q,t))^T$, we can write the dynamic equations for a single mode in matrix form as 
\begin{equation}  
\label{dyneq}
   B {\dot \wvec} = - A_a  \wvec + \vect{\eta}_{\mathrm{th}} + \vect{\eta}_{\mathrm{a}}
\end{equation}
where $B$ collects the transport coefficients: 
\begin{equation}  
        B = \left( \begin{array}{cc} 4 \eta q & 0 \\ 0 & 1/\left( \Omega  q^2\right) \end{array} \right) 
\end{equation} 
and $A_a$ is a matrix corresponding to elastic coefficients:
\begin{equation}  
        A_a = \left( \begin{array}{cc} \kappa q^4 + \sigma_{\rm eff} q^2 &  - \frac{1}{2} q^2 \left( \Lambda - F_{(2)} \right)  \\ - \frac{1}{2} q^2 \Lambda_a  & \chi_a \end{array} \right) 
\end{equation} 
with $\sigma_{\rm eff} = \sigma_0 + \sigma_{\rm dip}$. Note that for a lipid membrane the tension $\sigma_0$ will adjust when $\sigma_{\rm dip}$ changes to satisfy constraints on the total area of the membrane \cite{seifert95,lomholt11}.
The contributions of the thermal noise has been collected in the zero average noise vector $\vect{\eta}_{\mathrm{th}}$. The correlations of this vector with itself need to be  
\begin{equation} 
        \left\langle  \vect{\eta}_{\mathrm{th}}(q,t) \vect{\eta}_{\mathrm{th}}^\dagger(q,t') \right\rangle = 2 k_B T B L^2 \delta(t-t') 
\end{equation}
for the fluctuation dissipation theorem to be satisfied in equilibrium \cite{zwanzig01}. Here $\vect{\eta}_{\mathrm{th}}^\dagger$ is the transpose as well as complex conjugate of $\vect{\eta}_{\mathrm{th}}$. The term $\vect{\eta}_a$ contains the noise due to the fluctuations of the charge distribution inside the membrane   
\begin{equation} 
        \vect{\eta}_{\mathrm{a}} = -\frac{q^2}{2}\left( \begin{array}{c}  \bar{S}  \\  0 \end{array} \right)  
\end{equation} 
where $\bar{S}$ is the Fourier transform of the active noise $S$. We will assume that each pump operates independently with a typical relaxation time $\tau_p$. This leads to the correlations \cite{lomholt06b,prost98,lin06} 
\begin{equation} 
        \left\langle  \bar{S}(q,t) \bar{S}^*(q,t') \right\rangle = L^2n_\Sigma^0 \Gamma_a e^{- \frac{\left| t-t' \right| }{\tau_a} }  
\end{equation} 
where the noise strength $\Gamma_a$ will be of the order $(F_{(2)})^2$ and $\tau_a^{-1}=\tau_p^{-1} + D_{\rm sp} q^2$ with $D_{\rm sp}$ being the self-diffusion constant of a single pump. 

Let us now return to the issue of the influence of the pumping activity on $\Lambda_a$. In the absence of activity we will have $\Lambda_a=\Lambda$ and $A_a$ will be symmetric. This is no coincidence, since $A_a$ is then the Hessian matrix of second-order partial derivatives of the free energy $F$ with respect to ${\bar h}(q)$ and ${\bar n}_\Delta(q)$. The equality of the two off-diagonal elements of this Hessian matrix is thus an example of a Maxwell relation. We will argue that $A_a$ is also symmetric in the presence of activity, which means that $\Lambda_a=\Lambda-F_{(2)}$.
Our argument is that if we have a non-vanishing $F_{2}$ but $S(t)=0$, then this situation would correspond to having a different static distribution of charges in the active state of the pump relative to the passive one. But since there is no active noise in the system in this situation, then it would just correspond to a thermal system with different constraints on the position of the electric charges. 
Therefore the system can still be described by equilibrium statistical physics and consequentially also by a free energy depending on the constraints, with the corresponding Hessian matrix $A_a$ being symmetric. Since $A_a$ does not depend on dynamic noise
then this Maxwell relation must hold also when the noise $S(t)$ is turned on. This symmetrization of the dynamic equations, with activity modified coupling constant $\Lambda_a=\Lambda-F_{(2)}$ between pump diffusion and shape dynamics,
distinguishes the present model from previous work on force-multipoles \cite{manneville01,sankararaman02,lacoste05,lomholt06b}.

\begin{figure*}[t]
\begin{center}    
\begin{tabular}{c|c}
\includegraphics[scale=0.6]{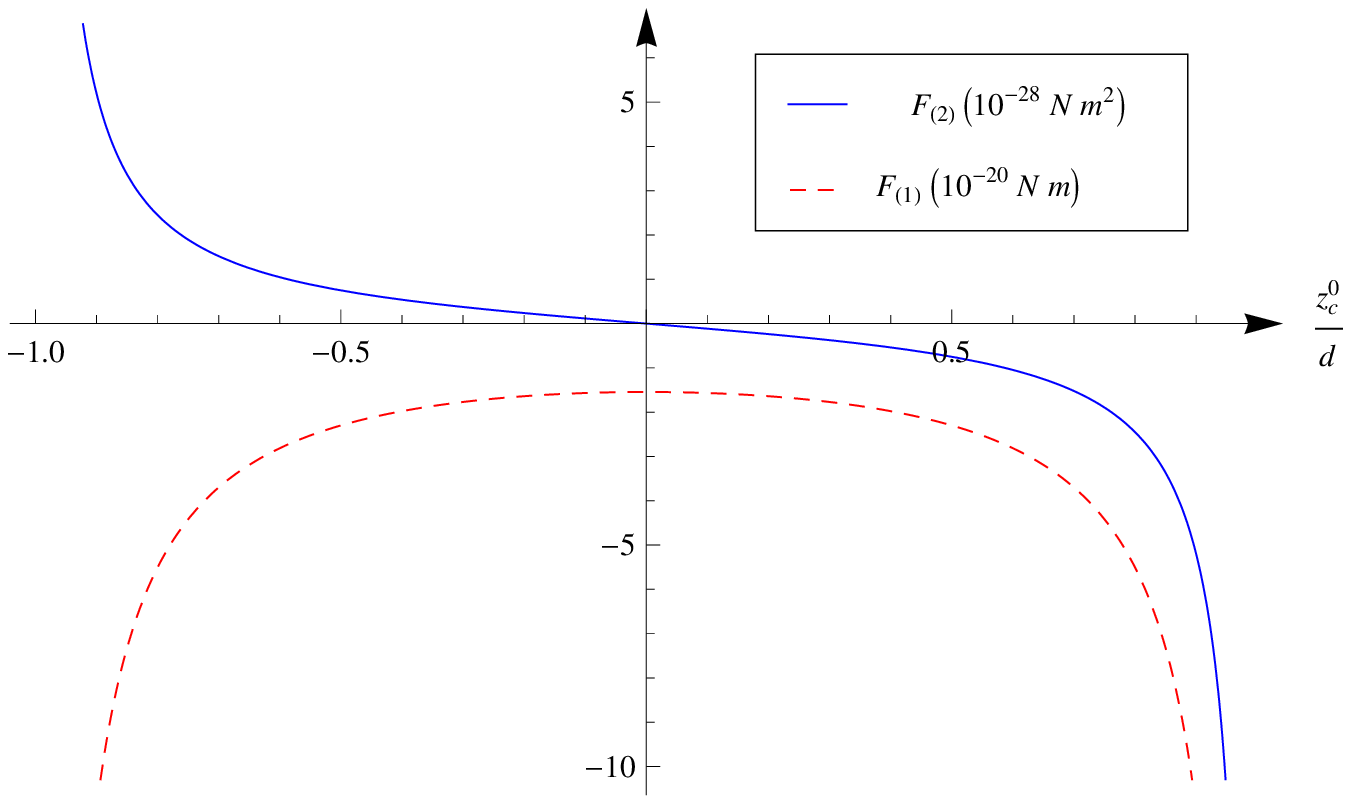} &
\includegraphics[scale=0.6]{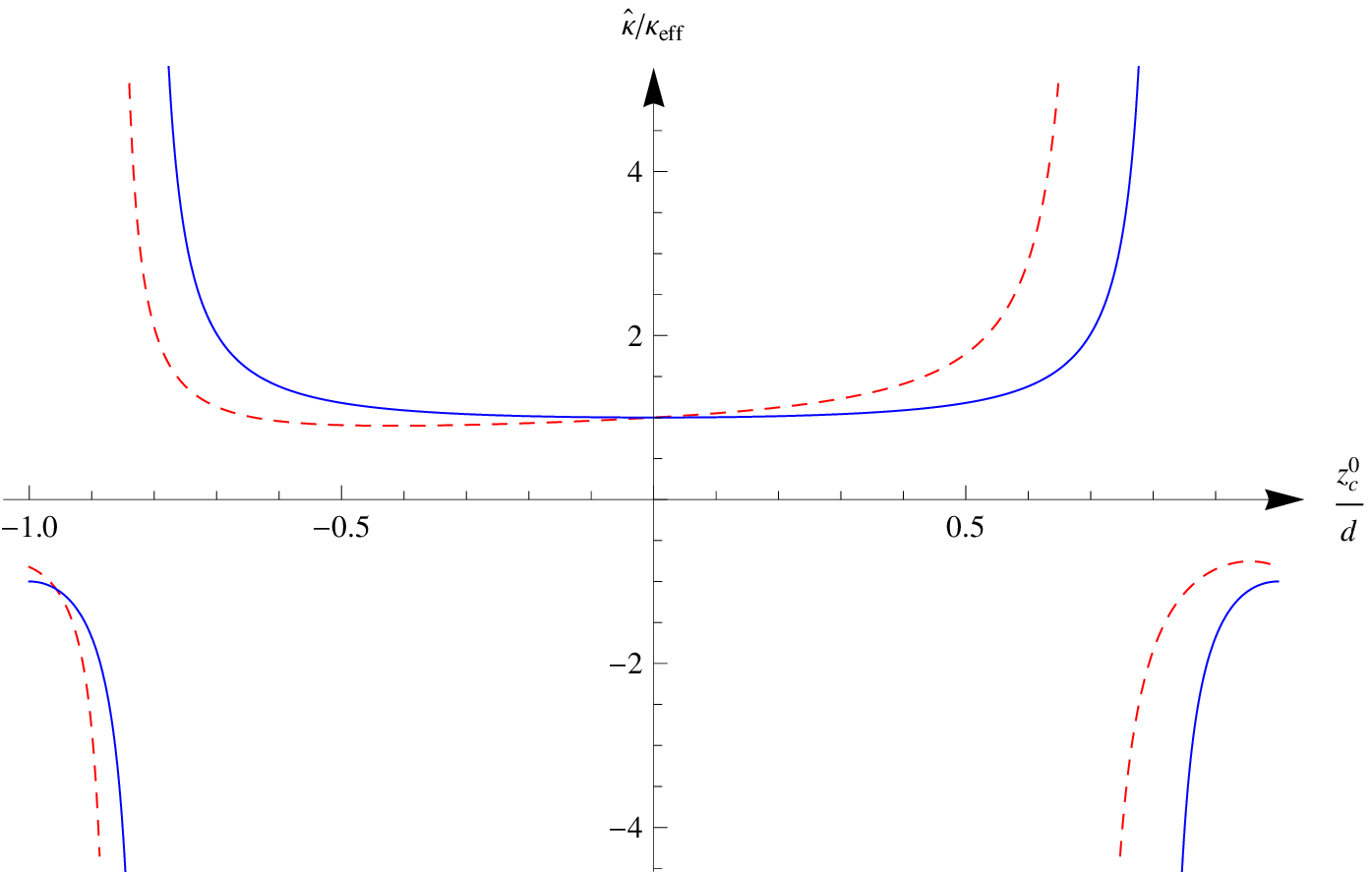}
\end{tabular}                             
\caption{The figure on the left is $F_{(2)}$ (continuous line, in units of $10^{-28}\,{\rm N}{\rm m}^2$) and $F_{(1)}$ (dashed line, in units of $10^{-20}\,{\rm Nm}$) as a function of the position of the charge $z_c^0$. The figure on the right is $\hat{\kappa}/\kappa_{\rm eff}$ for $\Lambda=0$ (continuous line) and $\Lambda=1.1\ 10^{-28} \ {\rm Jm}$ (dashed line) as a function of the position of the charge $z_c^0$. The parameters used for both figures are $n_\Sigma^0 = 10^{16}\,\mathrm{m^{-2}}$, $q_0 =e= 1.6\ 10^{-19} \, \mathrm{C}$, $2d = 5 \, {\rm nm}$, $\chi = \chi_a = k_B T/n_\Sigma^0$, $\kappa = 10 k_B T$, $\e = 80 \e_0$, $\e_2 = 2 \e_0$ and $\Gamma_a=(F_{(2)})^2$.}                                                                       
\label{tab:numapp}   
\end{center}      
\end{figure*}    

If we solve \Eqref{dyneq} for the equal time correlations of the shape $h(q,t)$ (using for instance the same methods as in \cite{lomholt06b}) we obtain
\begin{equation}
\label{fluctu} 
        \left\langle \left|\bar{h}(q)\right|^2  \right\rangle = \frac{k_B T L^2 }{\tilde{\kappa} q^4 + \sigma_{\rm eff} q^2} \left[ 1+ \frac{\Gamma_a n_{0,\Sigma}}{k_B T} \frac{q^3}{16 \eta} \frac{x_q}{\frac{1}{\tau_\kappa} + \frac{1}{\tau_D} } \right]
\end{equation}
where
\begin{equation}  
        \tilde{\kappa}\equiv \kappa - (\Lambda - F_{(2)})^2/(4\chi_a )
\end{equation}
is an effective active bending rigidity,
\begin{equation}
        x_q = \frac{\frac{1}{\tilde{\tau}_\kappa} + \frac{\tau_a}{\tau_D} \left( \frac{1}{\tau_D}+\frac{1}{\tau_\kappa} \right)+\frac{1}{\tau_D} }{\frac{1}{\tau_\kappa}+\frac{1}{\tau_D}+\frac{1}{\tau_a}+\frac{\tau_a}{\tau_D}\frac{1}{\tilde{\tau}_\kappa}}
\end{equation}
and we have introduced the inverse time-scales
\begin{align}
\tau_D^{-1} & = D_a q^2,\\
\tau_\kappa^{-1} & = (\kappa q^4 + \sigma_{\rm eff} q^2)/(4 \eta q),\\
\tilde{\tau}_\kappa^{-1} & = (\tilde{\kappa} q^4 + \sigma_{\rm eff} q^2)/(4 \eta q).
\end{align}
In the absence of active noise, $\Gamma_a=0$, then the quadrupole moment $F_{(2)}$ enters the fluctuations only through a modification of the bending rigidity. This deviates from the expressions given previously in \cite{manneville01,sankararaman02,lacoste05,lomholt06b} because the part of the dynamical equations corresponding to $A_a$ was not symmetrized in these papers. In the complete absence of activity ($F_{(1)} = F_{(2)} = \Gamma_a = 0$) we recover the equilibrium expression for the fluctuations
\begin{equation}
\label{eqfluctu}
        \left\langle \left|\bar{h}(q)\right|^2  \right\rangle_{\rm eq} = \frac{k_B T L^2 }{\hat{\kappa} q^4 + \sigma_0 q^2}
\end{equation}
where
\begin{equation}
        \hat{\kappa} = \kappa - \Lambda^2/(4\chi)
\end{equation}
is the effective passive bending rigidity.
If we take the limit of short wavelengths $q \rightarrow \infty$ in \Eqref{fluctu} assuming $D_{\rm sp}=D_a$ for simplicity we get
\begin{equation}
\label{qinffluc}
        \left\langle \left|\bar{h}(q)\right|^2 \right\rangle \approx  \frac{k_B T L^2}{{\tilde \kappa} q^4}\left({1 + \frac{ \Gamma_a n_\Sigma^0}{4 k_B T \kappa}}\right).
\end{equation}
In the opposite limit $q \rightarrow 0$ we find
\begin{equation}
\label{lowqlimit}
        \left\langle \left| \bar{h}(q) \right|^2 \right\rangle \approx  \frac{k_B T L^2}{\sigma_{\rm eff} q^2}.
\end{equation}
Thus the long wavelength behavior is completely controlled by tension also in the active case, while for the short wavelength the active noise (of strength $\Gamma_a$) increases fluctuations besides the influence of the asymmetry in average ion distribution in the pump. This asymmetry enters through the quadrupole moment $F_{(2)}$ of the active force, which together with $\Lambda$ couples fluctuations in pump density to membrane shape fluctuations. Note that the dipole moment $F_{(1)}$ only influences the shape fluctuations through a contribution to the tension $\sigma_{\rm eff}$. In between the short and long wavelength regimes there is a highly complex behavior influenced by for instance the pumping relaxation time $\tau_a$.

\section{Estimates of the electrostatic effect}

We cannot estimate the magnitude of the electrostatic effects by assuming a smeared distribution $\lambda(z_c)$ along the full width of the membrane. In this case for instance $F_{(1)}$ will become infinite due to diverging electrostatic attraction when the ion is close to the dielectric discontinuity at the water interfaces.
 Instead we will estimate the effects by assuming that the ion spends most of its time in the membrane around a fixed $z_c=z_c^0$ such that we have $\lambda(z_c)=\delta(z_c-z_c^0)$. In figure~\ref{tab:numapp} we show the dipole moment $F_{(1)}$, the quadrupole moment $F_{(2)}$ and $\hat{\kappa}/\kappa_{\rm eff}\equiv \lim_{q\to \infty}\langle |{\bar h}(q)|^2 \rangle / \langle |{\bar h}(q)|^2 \rangle_{\rm eq}$ as a function of $z_c^0$. The plots show that values of ${\hat \kappa}/\kappa_{\rm eff}\approx 2-3$ are attainable within our model. These values were obtained for a similar quantity called effective temperature (which also quantifies the increase in short wavelength fluctuations) for the pumps ${\rm Ca^{2+}}$-ATPase \cite{girard05} (in this reference they also measured $\Lambda =1.1\ 10^{-28} \ {\rm Jm}$) and bacteriorhodopsin \cite{manneville99} (here finding $\Lambda \approx 0$).
The effect at long wavelength (small $q$) is not immediately apparent from the values of $\sigma_{\rm dip}= n_\Sigma^0 F_{(1)}\lesssim -10^{-4} \ {\rm Nm^{-1}}$ obtained from figure~\ref{tab:numapp}. The values for $\sigma_{\rm dip}$ are large in magnitude compared to the tension change observed by a fluctuation analysis in \cite{faris09} upon pump activation.
However, as we have suggested in a previous paper \cite{lomholt11}, we will claim that
 a mechanism comes into play here that moderates the effect of such intrinsically applied tension:
 a negative contribution to the tension will make the membrane tend to expand. But due to the large compressional modulus for lipid membranes they can only expand very little before counteracting elastic forces set in. The situation studied here is more complicated than the one studied in \cite{lomholt11} though, since the tendency of a negative tension contribution to increase the area stored in long wavelength fluctuations is here competing for membrane area with the increase in short wavelength fluctuations (quantified by $\hat{\kappa}/\kappa_{\rm eff}$ in figure~\ref{tab:numapp}). We will study such effects of activity on tension in a future publication \cite{loubet11}.

\section{Conclusions}

In this letter we showed that the electrostatic interactions of an ion situated within a low dielectric lipid bilayer fits into the general model of membrane activity where the active pumps are modeled as fluctuating force-multipoles. This provide a possible microscopic picture behind this model. Contrary to the assumptions in previous work on the force-multipole model, the microscopic electrostatic picture presented here prescribes a symmetric coupling between the dynamics of the membrane shape and the density of ion pumps in the membrane.
 The effect of the mean quadrupole moment of the active force is thus simplified so that it acts as a coupling constant between the protein density and the membrane curvature.
This result might not hold for other microscopic mechanisms that generate force multipoles, e.g. steric interactions due to conformational changes of the pump during the active process.
The electrostatic model presented here relies on a number of simplifying assumptions. For instance it ignores interaction with free ions in the surrounding water, which must be present for the ion pumps to operate. However, the charge of the ions inside the membrane is already heavily screened by the large jump in dielectric constant at the membrane-water interface, so we do not expect this interaction to change the strengths of the involved forces by much.
For simplicity we have also assumed a symmetric distribution of pumps, i.e., that there are as many pumps pumping ions out of the vesicle as there are pumps pumping inward. As a consequence we do not take into account the possibility of a significant electric potential building up between the two sides of the membrane.
Such an electric potential would lead to additional effects by modifying the tension and bending rigidity \cite{lacoste07,ambjornsson07,lacoste09}.
Furthermore, the mechanical properties of the interior of the lipid membrane with pumps are assumed here as well as in previous work on force-multipoles \cite{manneville01,sankararaman02,lacoste05,lomholt06} to be similar to that of an incompressible fluid. And dielectric properties of the pumps are modelled as similar to the remaining lipid membrane. It would be interesting to test these last assumptions through for instance molecular dynamics simulations of lipid membranes with ion pumps studying electrostatic interactions in the system and the membranes mechanical response to such forces.

\section{Acknowledgments} We thank Per Lyngs Hansen and Himanshu Khandelia for helpful discussions. MEMPHYS - Center for Biomembrane physics is supported by the Danish National Research Foundation.

\bibliographystyle{eplbib}
\bibliography{refs}

\begin{thebibliography}{10}
\bibitem{REM1}
We have for simplicity neglected any curvature dependency of the pump
  efficiency. This could have been included as a term $F_{(2)}'n_\Sigma^0 H$ in
  the expression for $Q$.
\end{thebibliography}

\end{document}